\documentclass[aps,onecolumn,superscriptaddress,amsmath,showpacs,tightenlines]{revtex4-1}
\usepackage[utf8]{inputenc}
\usepackage{graphicx}
\usepackage{subfigure}
\usepackage{natbib}
\usepackage{epsfig}
\usepackage{amsfonts}
\usepackage{amsthm,amsmath,amssymb}
\usepackage{mathrsfs}
\usepackage{epstopdf}
\usepackage{xcolor}
\usepackage[toc,page,title,titletoc,header]{appendix}
\usepackage[linesnumbered,ruled,vlined]{algorithm2e}

\newcommand{\KwInput}[1]{\textbf{Input:} #1}
\newcommand{\KwOutput}[1]{\textbf{Output:} #1}

\begin{document}

\title{Dissipation-driven quantum generative adversarial networks}
\author{He Wang}
\affiliation{College of Physics, Jilin University,\\Changchun 130021, China}%
\affiliation{State Key Laboratory of Electroanalytical Chemistry, Changchun Institute of Applied Chemistry,\\Changchun 130021, China.}%
\author{Jin Wang}
\email{jin.wang.1@stonybrook.edu}
\affiliation{Department of Chemistry and of Physics and Astronomy, Stony Brook University, Stony Brook,\\NY 11794-3400, USA}%

\begin{abstract}

Quantum machine learning holds the promise of harnessing quantum advantage to achieve speedup beyond classical algorithms. Concurrently, research indicates that dissipation can serve as an effective resource in quantum computation. In this paper, we introduce a novel dissipation-driven quantum generative adversarial network (DQGAN) architecture specifically tailored for generating classical data. Our DQGAN comprises two interacting networks: a generative network and a discriminative network, both constructed from qubits. The classical data is encoded into the input qubits of the input layer via strong tailored dissipation processes. This encoding scheme enables us to extract both the generated data and the classification results by measuring the observables of the steady state of the output qubits. The network coupling weight, i.e., the strength of the interaction Hamiltonian between layers, is iteratively updated during the training process. This training procedure closely resembles the training of conventional generative adversarial networks (GANs). By alternately updating the two networks, we foster adversarial learning until the equilibrium point is reached. Our preliminary numerical test on a simplified instance of the task substantiate the feasibility of our DQGAN model.

\end{abstract}

\maketitle
\section{Introduction}

In recent years, machine learning has witnessed remarkable progress, transforming diverse domains from image recognition to autonomous vehicles. Among the prominent machine learning models that have captured significant attention are generative adversarial networks (GANs) \cite{IJG14}. GANs are specifically designed to generate novel samples that closely mimic real data by engaging a generator and a discriminator in a competitive machine-learning game. The generator of GANs synthesizes fabricated data samples within a defined domain, such as images, text, or audio. Its objective is to produce outputs that share statistical distributions akin to real data, aiming to mislead the discriminator. Conversely, the discriminator's role is to distinguish between the fake data samples crafted by the generator and genuine data instances. This adversarial interaction between the generator and discriminator forms the crux of the GAN framework. The overarching objective of this adversarial game is to achieve a Nash equilibrium, where the generator generates data that manifests genuine statistical properties, and the discriminator has a probability of correctly classifying real and fake samples of approximately  $\frac{1}{2}$. This equilibrium signifies the optimal state where the generator has successfully learned to generate data that is indistinguishable from real data, creating a compelling and realistic representation of the originating domain.

In parallel, the realm of quantum computing has witnessed remarkable advancement, achieving a pivotal milestone with experimental demonstrations of quantum supremacy \cite{MN10, GE19, AF19, HSZ20}. This accomplishment has been accompanied by the inception of groundbreaking algorithms such as Shor's algorithm \cite{PWS94}, Grover's search algorithm \cite{LKG96}, Harrow-Hassidim-Lloyd (HHL) algorithm \cite{AWH09}, and the quantum approximate optimization algorithm (QAOA) \cite{FE14}. These advancements have ignited fervent interest in exploring the potential quantum advantages over their classical counterparts, giving rise to quantum machine learning. The primary objective of quantum machine learning is to accelerate computations in the field of machine learning by harnessing the inherent quantum properties of entanglement and superposition \cite{ML23}. This pursuit has yielded notable theoretical breakthroughs and practical implementations, encompassing quantum principal component analysis \cite{LS14}, quantum support vector machines \cite{PR14}, variational quantum eigensolvers \cite{JRM16}, quantum Boltzmann machines \cite{MHA18}, quantum reinforcement learning \cite{DD08, GDP14, ML23b}, quantum feature spaces and kernels \cite{HV19}, quantum reservoir computing \cite{LCGG21}, and quantum generative adversarial networks (QGANs) \cite{SL18, PLDD18, LH19, CZ19, HZS20, BK21}, among others. In this paper, we delve specifically into the study of QGANs.

It is of paramount importance to acknowledge that all quantum systems unavoidably interact with their ambient environment, leading to dissipation and decoherence, which can detrimentally impact their quantum properties \cite{BHP06}. Mitigating the adverse effects of dissipation and decoherence is a fundamental challenge, particularly in the context of noisy intermediate-scale quantum (NISQ) devices \cite{JP18}. Intriguingly, it is crucial to recognize that the influence of the environment on quantum systems is not always detrimental and can, in certain scenarios, present opportunities for advantage. For instance, coupling two initially uncorrelated quantum detectors to a common field can induce quantum correlations between them \cite{FB05, BEG13, HW21}. Furthermore, under specific conditions, elevated temperatures can even amplify quantum correlations \cite{BEG13, HW21}. Remarkably,  dissipation can serve as a valuable resource for achieving universal quantum computation \cite{BK08, FV09}. As a consequence, the concept of dissipative engineering, which utilizes and exploits dissipation, is emerging as a powerful tool. It has enabled the preparation of intricate many-body states \cite{FV09, BK08} and facilitated the realization of exotic phases of matter \cite{SD11, HW23} and has been used to perform quantum machine learning tasks \cite{DT19, UK22, UK23, UK23b, GE21, GE22, GE23, GE23b, HWCB23}.

In our previous work, we explored the potential of a dissipative quantum classifier and showcased its remarkable performance across various classification tasks \cite{HWCB23}. This paper delves into the question of whether dissipation can serve as a computational resource for learning the distribution of classical data and generating novel data samples. Existing efforts in dissipative quantum generative adversarial networks (DQGANs) based on quantum circuits \cite{BK21} primarily employ dissipation as a byproduct of layer-to-layer transformation mappings, rather than a true computational resource that autonomously drives computation to completion. In contrast, this paper presents a genuine dissipative-driven quantum adversarial network. Current quantum generative networks can broadly be categorized into two types: those that generate classical data \cite{SL18, HZS20} and those that generate quantum data \cite{SL18, PLDD18, LH19, CZ19, BK21}. The model introduced in this paper primarily focuses on generating classical data. The input classical data is mapped to dissipative modes acting on input qubits, and the computational result is extracted from the steady state of output qubits through the measurement of certain observables. We demonstrate the effectiveness of our model in learning and generating classical data using a simplified example.

The remainder of this paper is organized as follows. Section \ref{model} introduces the model under investigation and presents numerical results for our DQGANs. Section \ref{summary} concludes with a discussion of the results and future directions.

\section{From GANs to DQGAN}\label{model}

\subsection{Classical GANs}

This subsection delves into a review of the classical GANs architecture. GANs employ adversarial training to ensure that the samples generated by the generative network adhere to the real data distribution. Within the framework of generative adversarial networks, two networks engage in an adversarial training process. The first, the discriminative network, aims to accurately determine whether a sample originates from the real data distribution $p_r(\pmb x)$ or is produced by the generative network $p_g(\pmb x)$.  As a binary classifier, the discriminator assigns a label $y=1$ to samples from the real distribution and $y=0$ to samples generated by the generative network. The discriminator's output $D(\pmb x;J_D)$  represents the probability that a sample belongs to the real data distribution, expressed as $p(y=1|\pmb x)=D(\pmb x;J_D)$.$J_D$ is the parameter matrix of the discriminative network. Conversely, the probability that a sample is generated by the generative network is $p(y=0|\pmb x)=1-D(\pmb x;J_D)$.

The second network, the generative network, strives to synthesize samples $G(\pmb z;J_G)$ that the discriminator cannot distinguish from real data samples. The input $\pmb z$ is drawn from a random distribution $p_{rand} (\pmb z)$. $J_G$ is the parameter matrix of the generative network. These two networks, with conflicting objectives, are iteratively trained in an alternating manner. Upon convergence, if the discriminator can no longer accurately classify the source of a sample, it implies that the generative network has successfully learned to produce samples that conform to the real data distribution.

\subsection{DQGANs}

\begin{figure}[!ht]
    \centering
\includegraphics[width=4in]{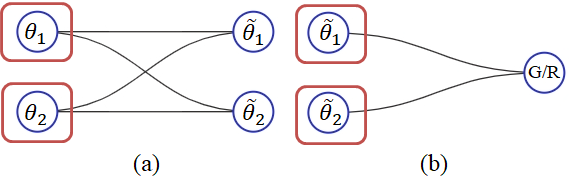}
\caption{\label{fig1} Figure \ref{fig1}(a,b) presents a simplified schematic of our DQGAN architecture. Both the generative network (a) and the discriminative network (b) incorporate interactions between the output qubits and the input qubits. Notably, the input qubits are subjected to engineered strong dissipative processes. The final computational outcome is extracted from the steady state of the output qubits.}
\end{figure}

This subsection introduces our proposed dissipative-driven quantum generative adversarial network (DQGAN) architecture. Figure \ref{fig1} presents a simplified illustration of the model. Both the generative network and the discriminative network in our DQGAN are composed of two qubit-based layers: an input layer and an output layer. Notably, there is no inter-qubit interaction within each layer. The coupling Hamiltonian between an output qubit and the input layer is defined as

\begin{equation}\label{hamiltonian}
\pmb H_{m} = \sum_n\vec{\pmb\sigma^n} \cdot (J_{nm} \vec{\pmb\sigma^{m}}),
\end{equation}

where the coupling matrix $J_{nm}$ describes the interaction between the $n$th input qubit and the $m$th output qubit. The input qubits are subjected to engineered strong dissipative processes, which can be described by a Lindblad master equation,

\begin{equation}
    \frac{\partial \pmb \rho(\tau) }{\partial \tau} = -i\left[\pmb H_m,\pmb \rho(\tau)\right] + \Gamma { \cal \pmb D} [\pmb \rho(\tau)] =-i\left[\pmb H_m,\pmb \rho(\tau)\right] + \Gamma \sum_{\alpha}\sum_{n=1}^N{ \cal\pmb D}_{\pmb L_n^\alpha} [\pmb \rho(\tau)],
\end{equation}

Here, $\pmb \rho$ represents the total density matrix encompassing both the input layer and the $m$-th output qubit. Additionally, the dissipator $\mathcal{\pmb D}_{\pmb L_n^\alpha}$ is introduced to account for the $\alpha$-th strong dissipative modes affecting the n-th input qubit. The dissipator adheres to the standard Lindblad form, defined as $\mathcal{\pmb D}_{\pmb L_n^\alpha} \pmb X =\pmb L_n^\alpha \pmb X \pmb L^{\alpha\dagger}_n - \frac{1}{2} (\pmb L^{\alpha\dagger}_n \pmb L_n^\alpha \pmb X + \pmb X \pmb L^{\alpha\dagger}_n\pmb L_n^\alpha)$. A schematic representation of the system is depicted in Fig.\ref{fig1}.

We choose the Lindblad operators as $\pmb L_n^1=\sqrt{\frac{1+\mu}{2}}|s^\perp_n\rangle\langle s_n|$, and $\pmb L_n^2 =\sqrt{\frac{1-\mu}{2}} |s_n\rangle\langle s^\perp_n|$. Here, $|s_n\rangle$ represents an arbitrary vector in the Hilbert space of the n-th qubit, satisfying the orthogonality condition $\langle s_n|s^\perp_n\rangle=0$. In quantum machine learning algorithms, efficiently encoding classical data into a quantum system or state is crucial for practical implementation. Conventionally, classical information is encoded into a quantum state through a parameterized quantum circuit. In this paper, however, we propose a novel approach where the classical data is encoded into the steady state of the input qubits through engineered dissipation \cite{HWCB23}. To achieve this, we parameterize the states $|s_n\rangle$ and $|s_n\rangle$ as follows,

\begin{equation}\label{evl}
|s_n\rangle=\left(
\begin{array}{c}
      \cos(\theta_n/2)
      \\
      \sin(\theta_n/2) 
\end{array}\right),\qquad |s_n^\perp\rangle=\left(
    \begin{array}{c}
      \sin(\theta_n/2) 
      \\
      -\cos(\theta_n/2)
    \end{array}
  \right).
\end{equation}
By tailoring dissipations on the n-th input qubit, we can encode the classical data $\theta_n$ into the system, as illustrated in Fig. \ref{fig1}. 

Assuming the dissipator $\mathcal{\pmb D}_{\pmb L_n^\alpha}$ is diagonalizable with a non-degenerate steady state, we define its eigenbasis $\{\pmb\psi_k^n\}$ with eigenvalues ${\xi_k^n}$ (where $\xi_0^n = 0$). Additionally, we introduce a complementary basis $\{\pmb\varphi_k^n\}$ satisfying the condition $Tr(\pmb\psi_l^n\pmb\varphi_k^m)=\delta_{mn}\delta_{lk}$. Consequently, the eigenbasis of the total dissipator can be expressed as $\pmb\Psi_{\{a,b...\alpha\}}=\pmb\psi_a^1\otimes\pmb\psi_b^2\dots\otimes\pmb\psi_\alpha^N$ with eigenvalue  $\Xi_{\{a,b...\alpha\}}=\xi_a^1+\xi_b^2+\dots\xi_k^n$. Similarly, we define the complementary basis as $\pmb\Phi_{\{a,b...\alpha\}}=\pmb\varphi_a^1\otimes\pmb\varphi_b^2\dots\otimes\pmb\varphi_\alpha^N$.

In the limit of strong dissipations ($\Gamma\rightarrow\infty$), the dynamics of the entire system become confined to a decoherence-free subspace spanned by $\pmb \psi_0^1 \otimes \pmb \psi_0^2\dots\otimes\pmb \psi_0^N\otimes \pmb \rho_m(\tau)$ after a short relaxation time. Tracing out the degrees of freedom of the input qubits, the evolution of the $m$th output qubit can be described by an effective master equation, as shown in previous works  \cite{EMK12,VP18,VP21},

\begin{equation}\label{effmaster}
    \frac {\partial \pmb \rho_m}{\partial \tau} = -i \left[\pmb h_D+ \tilde{\pmb H}_a/\Gamma,\pmb \rho_m \right]+\frac{1}{\Gamma}\tilde{\mathcal{\pmb D}}[\pmb \rho_m]
\end{equation}

where $\pmb h_D=Tr_{\bar{S}} \left( \left(\pmb\Psi_{\{0,0...0\}}\right)\pmb H_m \right)= Tr_{\bar{S}} \left( \left(\pmb \psi_0^1 \otimes \pmb \psi_0^2\dots\otimes\pmb \psi_0^N\right)\pmb H_m \right)=Tr_{\bar{0}} \left( \pmb\Psi_{\pmb 0}\pmb H_m \right)$. The super-operator $Tr_{\bar{S}}(\cdot)$ denotes the partial trace operation, eliminating all degrees of freedom except those belonging to the system. The super-index $\pmb 0$ signifies the specific configuration $\{0,0,\ldots,0\}$,  representing a unique state within the decoherence-free subspace. The complete form of the effective master equation (Eqn.~\ref{effmaster}) can be found in Appendix~\ref{app:A}. 


\section{Algorithm and numerical test}

In this section, we will introduce two different algorithms for training DQGANs.  In the first algorithm, we optimize a linear function of the expected value. The second algorithm is a modification of the original GANs algorithm, where we define a conditional probability by incorporating the expectation value of the output spin into a sigmoid function, and our optimization function is the log-likelihood.

\subsection{Algorithm 1}
Building upon the previous section, we proceed with training DQGANs based on the first algorithm.
In our approach to data generation, we leverage the dissipation mode associated with each random sample $\vec{\theta}_{rand}$ as input to the generative network. The generated data, $\theta^m_g=\arccos{\langle\pmb \sigma^m_z\rangle}$, is subsequently obtained by measuring the average value of the observable $\pmb \sigma_z$ at the steady state. We then feed both the generated samples $\vec{\theta}_g$ and real data samples $\vec{\theta}_r$ into the input layer of the discriminative network. To assess the likelihood of a sample originating from the real data distribution, we perform a projective measurement on the discriminative network's output, as expressed by the formula: 

\begin{equation}
p(\vec{\theta}\in\{\vec{\theta}_r\} )=tr(|R\rangle\langle R| \pmb \rho_{ssod}),
\end{equation}

where $\pmb \rho_{ssod}$ is the steady state of the output qubit of the discriminative network. The probability of the data being sampled from the generated data is then given by $p(\vec{\theta}\in\{\vec{\theta}_g\} )=1-p(\vec{\theta}\in\{\vec{\theta}_r\} )=tr(|F\rangle\langle F| \pmb \rho_{ssod})$. Here, $|F\rangle$ and $|R\rangle$ are arbitrary states that satisfy the orthonormal condition. Consequently, the optimization objective for DQGANs training can be formalized as the following adversarial task:

\begin{equation}
min_{J_G}max_{J_D} \frac{1}{S}\sum_{i=1}^{S}\langle R|\pmb \rho_{ssod}(\vec{\theta}^i_r;J_D)|R\rangle + \langle F|\pmb \rho_{ssod}(G(\vec{\theta}_{rand}^i;J_G);J_D)|F\rangle,
\end{equation}

The discriminator and generator networks are updated in an alternating fashion. During the discriminator update step, the objective is to maximize the following loss function: 

\begin{equation}\label{L_D1}
L_D= \frac{1}{S}\sum_{i=1}^{S}\langle R|\pmb \rho_{ssod}(\vec{\theta}^i_r;J_D)|R\rangle + \langle F|\pmb \rho_{ssod}(G(\vec{\theta}_{rand}^i;J_G);J_D)|F\rangle.
\end{equation}

In contrast, during the generator network update step, the objective is to maximize the following loss function:

\begin{equation}\label{L_G1}
L_G= \frac{1}{S}\sum_{i=1}^{S}\langle R|\pmb \rho_{ssod}(G(\vec{\theta}_{rand}^i;J_G);J_D)|R\rangle.
\end{equation}
The network weights are updated iteratively using the mini-batch gradient ascent algorithm. The first algorithm is shown below:
\begin{algorithm}
    \caption{The first algorithm of training of DQGAN}
    \label{alg:AOA}
    \SetAlgoNlRelativeSize{0}

    \quad\KwInput{The training set $\mathbb{D}$, the number of adversarial training iterations $T$, the number of mini-batch samples $S$.}
    \quad \quad\quad\KwOutput{The generative network outputs a new sample from a latent distribution.}
    
    \quad Initialize the coupling parameters of both the generative and discriminative networks, $J_G$ and $J_D$, to random values.
    
    \quad\For{$t \in \{1, 2, \dots, T\}$}{
       \quad \quad Train the discriminative network.
        
       \quad \quad Draw $S$ samples $\{\vec{\theta}_r^i\}$ from the training set $\mathbb{D}$. 
     \\\quad \quad Draw $S$ samples $\{\vec{\theta}_{rand}^i\}$ from the random distribution. 
        
      \quad  \quad Update $J_D$ using stochastic gradient ascent, the gradient is \\
      \quad \quad \quad  $\frac{\partial}{\partial J_G}[\frac{1}{S}\sum_{i=1}^{S}\langle R|\pmb \rho_{ssod}(\vec{\theta}^i_r;J_D)|R\rangle + \langle F|\pmb \rho_{ssod}(G(\vec{\theta}_{rand}^i;J_G);J_D)|F\rangle]$
        
       \quad \quad  Train the generative network. 
      \\\quad  \quad  Draw $S$ samples $\{\vec{\theta}_{rand}^m\}$ from the random distribution.   
        
       \quad \quad  Update $J_G$ using stochastic gradient ascent, the gradient is \\
     \quad  \quad \quad  $\frac{\partial}{\partial J_G}[\frac{1}{S}\sum_{i=1}^{S}\langle R|\pmb \rho_{ssod}(G(\vec{\theta}_{rand}^i;J_G);J_D)|R\rangle]$
    }
    
    \quad \Return The generative network.
\end{algorithm}

\begin{figure}[!ht]
    \centering
\includegraphics[width=1\textwidth,height=0.3\textwidth]{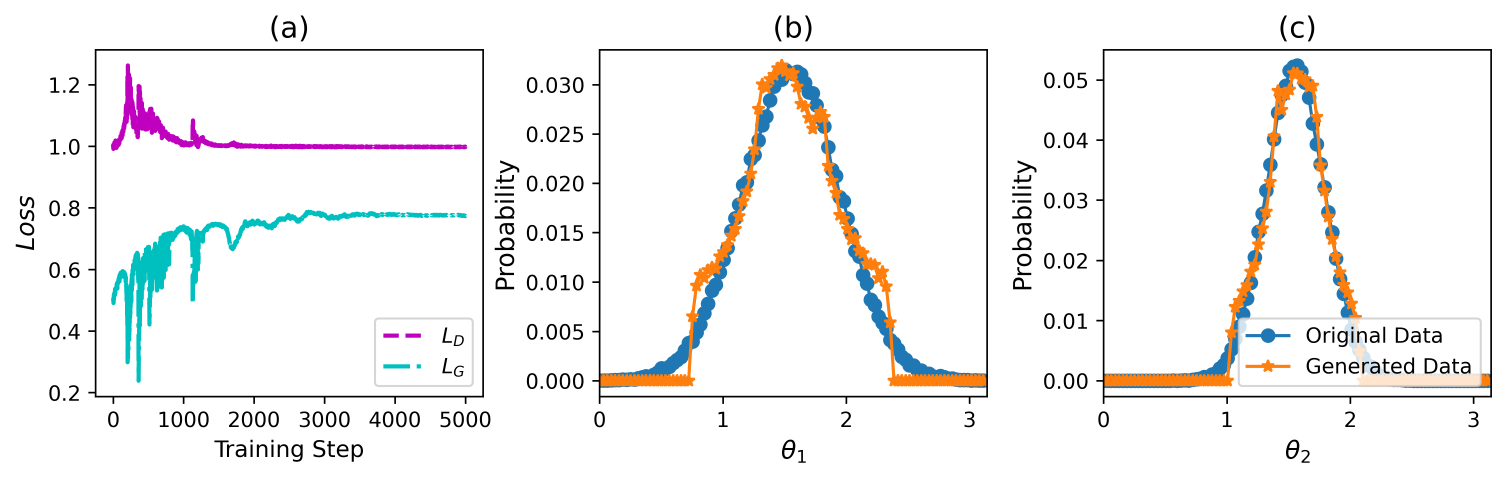}
\caption{\label{fig2}  Tracking of DQGANs. Fig.\ref{fig2}(a): The trends of two loss functions w.r.t. the training steps. During DQGANs training, the learning rate of cosine annealing is used, i.e., $\eta=\eta_{min}+\frac{1}{2}(\eta_{max}-\eta_{min})(1+cos(\frac{t}{T}\pi))$. $\eta_{max}=0.005$ and $\eta_{min}=0.0005$. Fig.\ref{fig2}(b,c) showcase the marginal probability distribution of $\theta_1$ and that of $\theta_2$ respectively. The real data is drawn from the set $\theta_2=a\theta_1+b+\mathscr{N}(0,0.01)$ with $p(\theta_1)=\mathscr{N}(\pi/2,0.4)$, $a=-0.6$ and $b=0.8\pi$.}
\end{figure}

We numerically tested DQGANs in this paper on a simple example. The real data is drawn from a linear relationship with Gaussian noise: $\theta_2=a\theta_1+b+\mathscr{N}(0,0.01)$ and $p(\theta_1)=\mathscr{N}(\pi/2,0.4)$. As input to the generative network, a pair of  $(\theta_1,\theta_2)$  is randomly chosen within the $\theta_1-\theta_2$ plane. The corresponding dissipative modes then act on the input qubits. After a short relaxation period, the generated data is obtained from the generative network's output layer, while the discriminative network provides classification results. During DQGAN training, a cosine annealing learning rate schedule was employed. Figure \ref{fig2}(a) illustrates the trends of the loss functions. Two functions engage in adversarial training by alternately increasing or decreasing, and after a certain number of training steps, the two networks reach a compromise, at which point each function converges to a stable value. By examining equation Eqn.\ref{L_D1}, we find that when the discriminative network is trained to its optimal state, $p(\vec{\theta}\in\{\vec{\theta}_g\} )=0.5$ and $p(\vec{\theta}\in\{\vec{\theta}_r\} )=0.5$, resulting in $L_D=1$ and $L_G=0.5$. This indicates that in the ideal state, the discriminative network can correctly classify real samples as real with a probability of $50\%$  and classify generated samples as fake with a probability of $50\%$.  However, in our results, $L_D\approx1$ and $L_G\approx0.77$, indicating a local optimum. Figure \ref{fig4}(a,b) depicts the probability distributions of real and generated data using the first algorithm, showing similar outlines. The marginal probability distributions in Figure \ref{fig2}(b,c) reveal high similarity, although a small portion of the real data distribution remains uncaptured by DQGAN. To quantify this similarity, we compute the Hellinger Distance, a measure of the difference between two probability distributions: $D_H=\frac{1}{\sqrt{2}}\sqrt{\sum_i^N (\sqrt{p_r^i}-\sqrt{p_g^i})^2}$. The obtained value of $D_H=0.137$ indicates that our DQGAN successfully captures the essential characteristics of the real data while introducing a degree of diversity and creativity into the generated samples.

\subsection{Algorithm 2}
In this subsection, we train the DQGANs according to the second algorithm. Identical to the prior setup, the generative network yields data samples denoted by $\theta^m_g=\arccos{\langle\pmb \sigma^m_z\rangle}$, where $\langle\pmb \sigma^m_z\rangle$ represents the local spin expectation value of the generative network. As before, both the generated data samples $\vec{\theta}_g$ and real data samples  $\vec{\theta}_r$ are fed into the discriminative network. The discriminative network's output, determined by measuring the average value of $\pmb \sigma_z$ in the steady state of its output layer, defines the probability that the input samples belong to the real data distribution,

\begin{equation}
D(\vec{\theta};J_D)=p(y=1|\vec{\theta})=\frac{1}{1+\exp{(-k\langle\pmb \sigma_z\rangle})},
\end{equation}

which corresponds to the Sigmoid function and $k$ is a control parameter. Accordingly, the discriminative network considers the probability of the input samples belonging to the generated data is $ p(y=0|\vec{\theta})=\frac{\exp{(-k\langle\pmb \sigma_z\rangle})}{1+\exp{(-k\langle\pmb \sigma_z\rangle})}$.

Mirroring the dynamics of classical GANs, our DQGANs engage in a minimax game. The discriminative network aims to maximize its ability to distinguish between real and generated samples, acting as the adversary in this competition. Consequently, its optimization objective seeks to maximize the following loss function,

\begin{equation}\label{lossD1}
max_{J_D} L_D= max_{J_D} \mathbb{E}_{\vec{\theta}\sim p(\vec{\theta})}[ylog(p(y=1|\vec{\theta}))+(1-y)log(p(y=0|\vec{\theta}))] ,
\end{equation}
Under the assumption that the input distribution $p(\vec{\theta})$ to the discriminative network is a uniform mixture of the real data distribution $p_r(\vec{\theta}_r)$ and the generated data distribution $p_g(\vec{\theta}_g)$, expressed as $p(\vec{\theta})=\frac{1}{2}(p_r(\vec{\theta}_r)+p_g(\vec{\theta}_g))$, the loss function in Eqn. \ref{lossD1} can be reformulated as follows:

\begin{equation}\label{lossD2}
max_{J_D} L_D= max_{J_D} \mathbb{E}_{\vec{\theta}_r\sim p_r(\vec{\theta}_r)}[log(D(\vec{\theta}_r;J_D))]+ \mathbb{E}_{\vec{\theta}_{rand}\sim p_{rand}(\vec{\theta}_{rand})}[log(1-D(G(\vec{\theta}_{rand};J_G);J_D))],
\end{equation}

In contrast to the discriminative network, the generative network seeks to deceive it. Its objective is to minimize the discriminative network's ability to distinguish between real and generated samples, thereby maximizing the probability of its own creations being classified as real. Consequently, the generative network's loss function is defined as:

\begin{equation}
max_{J_G} L_G= max_{J_G} \mathbb{E}_{\vec{\theta}_{rand}\sim p_{rand}(\vec{\theta}_{rand})}[log(D(G(\vec{\theta}_{rand};J_G);J_D))].
\end{equation}
The network weights are also updated iteratively using the mini-batch gradient ascent algorithm. The second DQGAN algorithm is shown below:

\begin{algorithm}
    \caption{The second algorithm of training of DQGAN}
    \label{alg:AOA}
    \SetAlgoNlRelativeSize{0}

    \quad\KwInput{The training set $\mathbb{D}$, the number of adversarial training iterations $T$, the number of mini-batch samples $M$.}
    \quad \quad\quad\KwOutput{The generative network outputs a new sample from a latent distribution.}
    
    \quad Initialize the coupling parameters of both the generative and discriminative networks, $J_G$ and $J_D$, to random values.
    
    \quad\For{$t \in \{1, 2, \dots, T\}$}{
       \quad \quad Train the discriminative network.
        
       \quad \quad Draw $M$ samples $\{\vec{\theta}_r^m\}$ from the training set $\mathbb{D}$. 
     \\\quad \quad Draw $M$ samples $\{\vec{\theta}_{rand}^m\}$ from the random distribution. 
        
      \quad  \quad Update $J_D$ using stochastic gradient ascent, the gradient is \\
      \quad \quad \quad  $\frac{\partial}{\partial J_D}[\frac{1}{M}\sum_{1}^{M}(log (D(\vec{\theta}_r^m;J_D)) +log (1-D(G(\vec{\theta}_{rand}^m;J_G);J_D)))]$
        
       \quad \quad  Train the generative network. 
      \\\quad  \quad  Draw $M$ samples $\{\vec{\theta}_{rand}^m\}$ from the random distribution.   
        
       \quad \quad  Update $J_G$ using stochastic gradient ascent, the gradient is \\
     \quad  \quad \quad  $\frac{\partial}{\partial J_G}[\frac{1}{M}log (D(G(\{\vec{\theta}_{rand}^m\};J_G);J_D))]$
    }
    
    \quad \Return The generative network.
\end{algorithm}

\begin{figure}[!ht]
    \centering
\includegraphics[width=1\textwidth,height=0.3\textwidth]{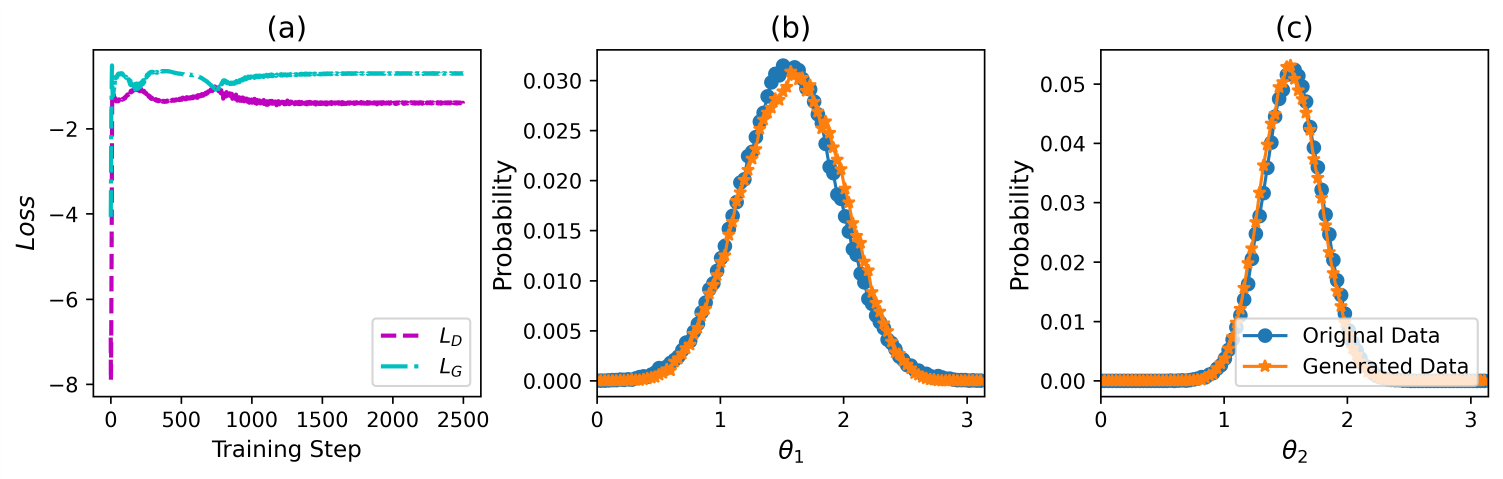}
\caption{\label{fig3}  Tracking of DQGANs. Fig.\ref{fig3}(a): The trends of loss functions w.r.t. the training steps. During DQGANs training, the learning rate of cosine annealing is used, i.e., $\eta=\eta_{min}+\frac{1}{2}(\eta_{max}-\eta_{min})(1+cos(\frac{t}{T}\pi))$. $\eta_{max}=0.005$ and $\eta_{min}=0.0001$. Fig.\ref{fig3}(b,c) showcase the marginal probability distribution of $\theta_1$ and that of $\theta_2$ respectively. The real data is drawn from the same set in Fig.\ref{fig2}.}
\end{figure}

\begin{figure}[!ht]
    \centering
\includegraphics[width=1\textwidth,height=0.3\textwidth]{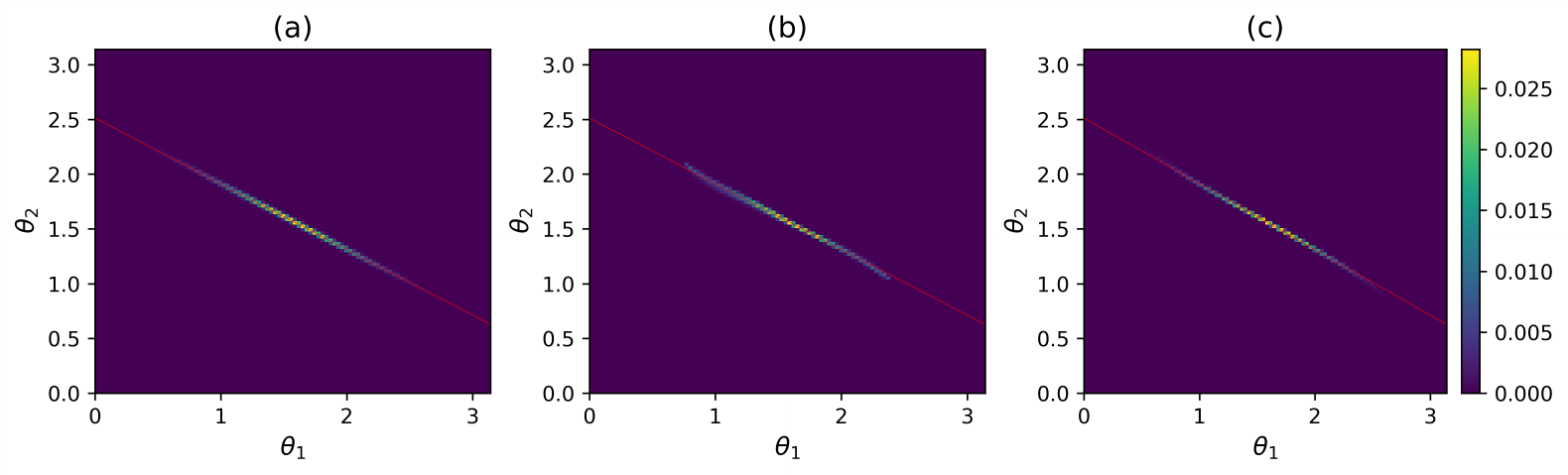}
\caption{\label{fig4}  The probability distribution of (a) real data and (b,c) generated data using various algorithms.}
\end{figure}

We are now training the network on the same instance using Algorithm 2. Figure \ref{fig3}(a) reveals the trends of the loss functions. Initially, both losses increase, reflecting the initial learning phase. Following this, an adversarial "tug-of-war" emerges, where one loss function increases while the other decreases. This dynamic continues until both losses converge to stable values. Notably, the loss function of the discriminative network $L_D$ converges to  $-2log2$ while $L_G$ converges to $-log2$. This aligns with the optimal loss function value for the discriminator at the equilibrium point in classical GANs \cite{IJG14}. The convergence arises from the close parallels between our DQGAN setting and the classical GAN framework, despite the fundamentally quantum nature of the underlying computation.  Figures \ref{fig3}(b) and (c) depict the marginal probability distributions of the real and generated data, respectively. Strikingly, they exhibit a close resemblance in their overall shape. This observation is further substantiated by the similar shapes evident in Figure \ref{fig4}(a,c). To quantify this similarity, we compute the Hellinger Distance, yielding $D_H=0.094$.  This value indicates that our DQGAN successfully captures the essential characteristics of the real data while introducing a degree of diversity and creativity into the generated samples.  Compared to Algorithm 1, this approach exhibits faster convergence and higher accuracy. However, it is crucial to note that a single example cannot conclusively demonstrate the absence of advantages for Algorithm 1; it may offer significant benefits in other applications.

\section{Summary}\label{summary}

In summary, we introduce and explore dissipative-driven quantum generative adversarial networks (DQGANs), a novel architecture where dissipation plays a dual role: encoding classical information into quantum steady states and acting as an internal driving force for computation. We demonstrate the feasibility of our approach through a simple example, showcasing the DQGAN's ability to learn and generate data. However, we acknowledge that this is a preliminary step, and several limitations require further development for the DQGAN to become a mature and reliable generative model. For instance, its vulnerability to mode collapse when dealing with complex data remains a challenge, mirroring known shortcomings of classical and quantum GANs. Fortunately, advanced algorithms like Wasserstein GANs (WGANs) \cite{MA17} and their quantum counterparts \cite{SC19} have proven effective in mitigating these issues. Addressing these limitations and refining the DQGAN architecture will be the focus of our future research.

\appendix
\section{The derivation of the effective master equation}\label{app:A}
Using the eigenbasis, we can decompose the total Hamiltonian as

\begin{equation}\begin{split}
    \pmb H_m &= \sum_{\{a,b,...\alpha\}} \pmb\Phi_{\{a,b...\alpha\}}\pmb g_{\{a,b,...\alpha\}}=\sum_{\{a,b,...\alpha\}} \pmb\Phi_{\{a,b...\alpha\}}^\dagger \otimes \pmb g_{\{a,b,...\alpha\}}^\dagger,\\  
\pmb g_{\{a,b,...\alpha\}} &= Tr_{\bar{0}} ((\pmb I\otimes\pmb\Psi_{\{a,b...\alpha\}}  )\pmb H_m).
\end{split}\end{equation}
By introducing the orthogonal vectors  $\vec{v}_n(\theta_n,\phi_n) = (\sin \theta_n \cos \phi_n, \sin \theta_n \sin \phi_n, \cos\theta_n )$, $\vec{v}_n' = \vec{v}_n \left(
  \frac{\pi}{2}-\theta_n,\phi_n+\pi \right)$, $\vec{v}_n'' = \vec{v}_n \left(\frac{\pi}{2},\phi_n+\frac{\pi}{2} \right)$, we obtain the following expression, 
  
\begin{equation}\begin{split}
\pmb g_{\pmb 0_n}&=\mu(\pmb J_n\vec{v}_n) \cdot \vec{\pmb\sigma}^m\\
\pmb g_{\pmb 1_n}&=(\pmb J_n\vec{v}_n') \cdot \vec{\pmb \sigma}^m - i (\pmb J_n\vec{v}_n'') \cdot\vec{\pmb \sigma}^m,\\
\pmb g_{\pmb 2_n}&=\pmb g_{\pmb 1_n}^\dagger\\
 \pmb g_{\pmb 3_n}&= 2 (\pmb J_n\vec{v}_n) \cdot \vec{\pmb\sigma}^m.
\end{split}\end{equation}
Define the coefficient matrix 
\begin{equation}\begin{split}
C_{\{a,b...\alpha\},\{a',b'...\alpha'\}}&= Tr \left( \pmb\Phi_{\{a,b...\alpha\}}^\dagger \pmb\Phi_{\{a',b'...\alpha'\}} \pmb\Psi_{\pmb 0} \right),
\\ -C_{\{a,b...\alpha\},\{a',b'...\alpha'\}}/\Xi_{\{a,b...\alpha\}}^*&=Y_{\{a,b...\alpha\},\{a',b'...\alpha'\}}=A_{\{a,b...\alpha\},\{a',b'...\alpha'\}}/2 +i B_{\{a,b...\alpha\},\{a',b'...\alpha'\}},
\end{split}\end{equation}
where $A_{\{a,b...\alpha\},\{a',b'...\alpha'\}}=Y_{\{a,b...\alpha\},\{a',b'...\alpha'\}}+Y_{\{a,b...\alpha\},\{a',b'...\alpha'\}}^*$ is a positive matrix and $B_{\{a,b...\alpha\},\{a',b'...\alpha'\}}=(Y_{\{a,b...\alpha\},\{a',b'...\alpha'\}}-Y_{\{a,b...\alpha\},\{a',b'...\alpha'\}}^*)/(2i)$ is a Hermitian matrix. We can derive

\begin{equation}\begin{split}
\tilde{\pmb H}_a &=  \sum_{Re(\Xi_{\{a,b...\alpha\}})<0,Re(\Xi_{\{a',b'...\alpha'\}})<0}  B_{\{a,b...\alpha\},\{a',b'...\alpha'\}}\pmb g_{\{a,b...\alpha\}}^\dagger \pmb g_{\{a',b'...\alpha'\}},\\
 \tilde{\mathcal{\pmb D}} \pmb R &= \sum_{Re(\Xi_{\{a,b...\alpha\}})<0,Re(\Xi_{\{a',b'...\alpha'\}})<0} A_{\{a,b...\alpha\},\{a',b'...\alpha'\}} \left( \pmb g_{\{a',b'...\alpha'\}} \pmb R \pmb g_{\{a,b...\alpha\}}^\dagger - \frac{1}{2} \{\pmb g_{\{a,b...\alpha\}}^\dagger \pmb g_{\{a',b'...\alpha'\}}, \pmb R\} \right)
\end{split}\end{equation}

It is crucial to emphasize that the validity of the aforementioned effective master equation hinges on the assumption of strong dissipations acting on the input qubits. While increasing the strength of these dissipations diminishes their direct impact on the output qubit, disregarding their influence altogether would be inaccurate. Notably, even under strong input dissipation, non-negligible dissipations may still affect the output qubit, and these must be considered to accurately predict the system's final steady state \cite{EMK12,VP18,VP21}.

Due to the locality of the Hamiltonian, the terms $\pmb g_{\{a,b,...\alpha...\beta...\}} = Tr_{\bar{0}} ((\pmb I\otimes\pmb\Psi_{\{a,b...\alpha...\beta...\}}  )\pmb H_m)=0$ vanish whenever any two indices $\alpha\geq0$ and $\beta\geq0$ are non-zero. Consequently, only the terms $\pmb g_{\pmb k_n} = Tr_{\bar{0}} ((\pmb I\otimes\pmb\Psi_{\{0,0...k...0\}}  )\pmb H_m)$ may not be zero. Here, the super-index $\pmb k_n$, denotes that only the n-th element is $k$ while the remaining elements in the set of N elements are zero. 

The eigenbasis and corresponding eigenvalues of the superoperator $\mathcal{\pmb D}_{\pmb L_n}$ are as follows,

\begin{equation}\begin{split}
\pmb\psi_0^n &= \frac{1+ \mu}{2}|s_n\rangle\langle s_n| + \frac{1- \mu}{2}|s_n^\perp\rangle\langle s_n^\perp|, \qquad  \xi_0^n=0,   \\
 \pmb\psi_1^n &= |s_n\rangle \langle s_n^\perp|, \qquad  \xi_1^n=-\frac{1}{2},\\
 \pmb\psi_2^n &= |s_n^\perp\rangle \langle s_n|, \qquad \xi_2^n=-\frac{1}{2},\\
 \pmb\psi_3^n &= |s_n\rangle \langle s_n|-|s_n^\perp\rangle \langle s_n^\perp|, \qquad \xi_3^n=-1,
\end{split}\end{equation}

and the complimentary basis are

\begin{equation}\begin{split}
\pmb\varphi_0^n&= \pmb I,\\
  \pmb\varphi_1^n&= |s_n^\perp\rangle \langle s_n|,\\
  \pmb\varphi_2^n&= |s_n\rangle \langle s_n^\perp|,\\
  \pmb\varphi_3^n&= \frac{1- \mu}{2} |s_n\rangle \langle s_n| - \frac{1+\mu}{2}|s_n^\perp\rangle \langle s_n^\perp|.
\end{split}\end{equation}  
 
We can now explicitly compute the coefficient matrix $C_{\{a,b...\alpha\},\{a',b'...\alpha'\}}= Tr \left( \pmb\Phi_{\{a,b...\alpha\}}^\dagger \pmb\Phi_{\{a',b'...\alpha'\}} \pmb\Psi_{\pmb 0} \right)$. The resulting coefficients are as follows: $C_{\pmb 0_m,\pmb 0_n}=\delta_{mn}$, $C_{\pmb 1_m,\pmb 1_n}=\delta_{mn}(1+\mu)/2$, $C_{\pmb 2_m,\pmb 2_n}=\delta_{mn}(1-\mu)/2$, $C_{\pmb 3_m,\pmb 3_n}=\delta_{mn}(1-\mu^2)/4$. All other coefficients are zero. From this, we can easily derive the matrices $A$ and $B$. Finally, the effective master equation for the output qubit is obtained,

\begin{equation}\begin{split}
\tilde{\pmb H}_a &=  0,\\
 \tilde{\mathcal{\pmb D}} \pmb R &= 2(1+\mu)\sum_{n} \left( \pmb g_{\pmb 1_n} \pmb R \pmb g_{\pmb 1_n}^\dagger - \frac{1}{2} \pmb g_{\pmb 1_n}^\dagger \pmb g_{\pmb 1_n} \pmb R-\frac{1}{2} \pmb R \pmb g_{\pmb 1_n}^\dagger \pmb g_{\pmb 1_n} \right)+2(1-\mu)\sum_{n} \left( \pmb g_{\pmb 1_n}^\dagger \pmb R \pmb g_{\pmb 1_n} - \frac{1}{2} \pmb g_{\pmb 1_n} \pmb g_{\pmb 1_n}^\dagger \pmb R-\frac{1}{2} \pmb R \pmb g_{\pmb 1_n} \pmb g_{\pmb 1_n}^\dagger \right)+\\&\quad+(1-\mu^2)/2\sum_{n} \left( \pmb g_{\pmb 3_n} \pmb R \pmb g_{\pmb 3_n}^\dagger - \frac{1}{2} \pmb g_{\pmb 3_n}^\dagger \pmb g_{\pmb 3_n} \pmb R-\frac{1}{2} \pmb R \pmb g_{\pmb 3_n}^\dagger \pmb g_{\pmb 3_n} \right),\\
\pmb h_{D} &= \sum_{n=1}^{N} \pmb g_{\pmb 0_n}.
\end{split}\end{equation}

\end{document}